\documentclass[12pt]{iopart}

\usepackage{epsf}
\usepackage{graphicx}
\usepackage{dcolumn}
\usepackage{bm}
\usepackage{ulem}

\begin{document}


\newcommand{\beq}{\begin{equation}}
\newcommand{\dd}{\partial}
\newcommand{\eeq}{\end{equation}}
\newcommand{\bea}{\begin{eqnarray}}
\newcommand{\eea}{\end{eqnarray}}



%

\begin{flushright}
\begin{tabular}{l}
UCLA/05/TEP/21
\end{tabular}
\end{flushright}

\title{Astrophysical bounds on 
supersymmetric dark-matter Q-balls
}

\author{Alexander Kusenko$^1$, Lee C. Loveridge$^1$, and \\
Mikhail Shaposhnikov$^2$
}
\address{$^1$Department of Physics and Astronomy, UCLA, Los Angeles, CA
90095-1547 \\ 
$^2$ Institute of Theoretical Physics, Ecole Polytechnique F\'ed\'erale de
    Lausanne, SB/ITP/LPPC, BSP 720,
    CH-1015, Lausanne, Switzerland}


\begin{abstract}
\vspace{0.1cm}

Stable baryonic Q-balls, which appear in supersymmetric extensions of
the Standard Model, could form at the end of cosmological inflation
from fragmentation of the Affleck -- Dine condensate.   We
reconsider  astrophysical constraints on such Q-balls as dark matter
candidates.  Baryonic Q-balls interact with matter by absorbing the
baryon number and, effectively, leading to a rapid baryon number
non-conservation.  We have recently shown that this process can occur
at a much faster rate than that used in previous calculations.   As a
consequence, stability of neutron stars imposes a stringent
constraint on the types of Q-balls that can be dark matter.  Only the
Q-balls that correspond to baryonic flat directions lifted by baryon-number
violating operators are allowed as dark-matter candidates.  

\end{abstract}


\maketitle

\renewcommand{\thefootnote}{\arabic{footnote}}
\setcounter{footnote}{0}

\section{Introduction}
Supersymmetry (SUSY) provides two types of dark matter candidates:
the lightest supersymmetric particle ({\it e.g.}, neutralino or
gravitino), and stable Q-balls.  SUSY Q-balls are nontopological
solitons that carry baryon number~\cite{q,nts_review,ak_mssm}.  They
are either stable or have lifetimes in excess of the age of the
universe in theories with gauge-mediated SUSY breaking~\cite{dks}. 
At the end of inflation, large SUSY Q-balls can form via
fragmentation of the Affleck-Dine
condensate~\cite{ad,dk,enqvist_review}.

SUSY Q-balls can interact with matter fermions.  A quark scattering
off a baryonic Q-ball can convert into an antiquark, while increasing
the baryon number of the Q-ball.  Hence, Q-balls catalyze baryon
number non-conservation in the sector of ordinary, non-supersymmetric
particles.  The  effective B-nonconservation catalyzed by Q-balls was
considered in Ref.~\cite{kkst} in connection with astrophysical
limits on Q-ball dark matter.  However, it was recently
shown~\cite{qi} that the rate of this process is considerably higher
than that used in Ref.~\cite{kkst}.  In this paper we reconsider 
astrophysical limits on dark matter Q-balls and show that a wide
variety of them are ruled out because they would destabilize neutron
stars on cosmologically short time scales.

\section{Interactions of SUSY Q-balls with matter}
In a previous paper ~\cite{qi} we reanalyzed the interaction of
Q-balls with ordinary matter. We found that quarks falling on a
Q-ball are reflected as antiquarks with a probability of  order one,
practically independent of the parameters of the theory. In other
words, Q-balls convert the matter into antimatter on their surface
(or antimatter into matter, if placed in an anti-matter environment).
Baryon number is conserved during this process: after reflection of
an antinucleon, the baryonic charge of the Q-ball increases by 2
units.

Having described the interaction of SUSY Q-balls with nucleons, we
now reexamine some bounds on dark-matter Q-balls.  The laboratory
limits do not change significantly from the earlier
estimates~\cite{kkst}.  This is because nucleons reflected from the
Q-ball surface as antinucleons have a very small mean free path in
the surrounding matter. They annihilate quickly, mainly into pions. The
energy release is somewhat higher than it was thought to
be~\cite{kkst}, but this is an order one effect, which does not make
a difference for most bounds. 

However, the analysis of astrophysical bounds, in particular those
from Q-balls captured in the neutron stars, must be done
differently.  The  rate of baryon number absorption by a Q-ball was
underestimated in Ref.~\cite{sw} by several orders of magnitude. One
can  expect a dramatic change in the astrophysical bounds because of
this. In fact, such rapid absorption of nucleons requires that we
reexamine the inner fluid dynamics of the neutron star.  Although the
dynamics of nuclear matter is difficult to describe, we understand it
well enough to know that certain types of Q-balls would consume a
neutron star in a few thousand years.  This is far less than the
lower bound lifetime of neutron stars and suggests that such Q-balls
must not exist in large amounts in the universe.

We will also address another important issue overlooked in the
earlier work.  Non-renormalizable operators alter the flat directions
in the potential and can modify the shapes of solutions for SUSY
Q-balls.  We will consider interactions of Q-balls of different
types.  

\section{SUSY Q-balls and the flat directions}
Stable Q-balls develop a VEV along some flat direction.  This flat
direction can be parameterized by a scalar field $\varphi$.  The VEVs
of squarks and sleptons are proportional to $\varphi$.  In theories
with gauge-mediated breaking of supersymmetry, the effective
potential along this flat direction is nearly constant, $V(\varphi)
\propto \Lambda^4$, up to some  $\varphi_{\rm max}$ which depends on
the type of the flat direction \footnote{We omit possible $\log$
factors, because they are not essential for the discussion.}.  There
exists a classical spherically symmetric solution of the field
equations, a non-topological soliton, in which 
\beq
\varphi = \phi_0 f(r)e^{i \omega t},  
\eeq 
where $f(r) \simeq \sin(\omega r)/(\omega r)$ for $r \leq
R=\pi/\omega$ and zero for $r>R$. It carries the global charge $Q$. 
If $\phi_0<\varphi_{\rm max}$, different parameters depend on $Q$ as
follows: 
\beq 
\phi_0 \sim \Lambda Q^{1/4},~~M \sim \Lambda Q^{3/4},~~
\omega \sim \Lambda Q^{-1/4}~.  
\label{fdprop}
\eeq 
We shall call the non-topological solitons of this type flat direction (FD) 
Q-balls.

Flat directions are in general lifted by higher dimension
non-renormalizable operators \cite{Gherghetta:1995dv} and the
effective potential starts to grow at $\varphi > \varphi_{\rm max}$.
Thus, when the charge $Q$ exceeds some critical value $Q_{c}$,
the properties of Q-balls change. The details of this analysis will
appear elsewhere \cite{ru}, here we reproduce the essentials only. 

Generally, the lifting terms can be written in the form
\beq
V^n(\phi)_{\mbox{\tiny lifting}}\approx  
\lambda_n M^4\left(\frac{\phi}{M}\right)^{n-1+m}
\left(\frac{\phi^*}{M}\right)^{n-1-m},
\label{lift}
\eeq
where the smallest possible $n$ and corresponding to it $m$ are some
integers that are determined by the structure of the flat direction
and can be found from  Tables 4 and 5 of ref.
\cite{Gherghetta:1995dv}. We will assume that $\lambda_n$ is of the
order one and that $M$ is the grand unification scale $\sim
10^{16}$GeV and that $\Lambda \sim 10^3$ GeV.

The relations (\ref{fdprop}) are valid only if the lifting potential
is small compared with $\Lambda^4$. So, the Q-ball properties must be
re-analyzed when their charge reaches the critical value $Q_c$
determined from the condition $V^n(|\phi|)_{\mbox{\tiny
lifting}}\simeq \Lambda^4$, and equal to
\begin{equation}
 Q_c\simeq \lambda_n^{-\frac{2}{n-1}}
\left(\frac{M}{\Lambda}\right)^{\frac{4n-12}{n-1}},
\label{b1}
\end{equation}
It can be seen  from \cite{Gherghetta:1995dv} that most flat
directions are lifted by monomials of dimension $n=4$.  At this low
suppression, only $Q$-balls with small charge $Q < Q_c \simeq
10^{17}$ can retain their $\propto Q^{3/4}$ energy dependence despite
the lifting. On the other end, {\it all} flat directions are lifted
by monomials of dimension $n=7$, which corresponds to a critical
charge  $Q_c \simeq 10^{34}$.

The fate of Q-balls with charges larger than $Q_c$ depends
crucially on the value of the integer $m$ in (\ref{lift}). For $m \neq 0$
the lifting potential not only breaks the degeneracy of the flat
direction, but introduces an explicit breaking of the global U(1)
symmetry responsible for Q-ball existence. For these flat directions
non-renormalizable operators induce strong baryon number
non-conservation,  which leads to fast Q-ball decay. As the value of
$Q_c$ happens to be much smaller than the baryon number of a
star, no astrophysical limits on this type of Q-balls can be derived.
These Q-balls absorb only a small fraction of the baryon material of a
star and disappear afterwords. In notations of
ref.~\cite{Gherghetta:1995dv}  Q-balls related to flat directions 
$QLe,~ QLd,~ Lude,~ QLde,~QLud,~QLude$ are of this type.

If, on the contrary,  $m=0$, the lifting term does not break the
global U(1) symmetry and the Q-ball charge can grow beyond $Q_c$, but
the  dependence of the parameters of the Q-ball on Q is modified.
Now, with the lifting terms incorporated, the quantity
$V(\varphi)/|\varphi|^2$ does have a minimum at non-zero finite value
of $\varphi$, and the analysis of \cite{q} is applicable. Thus, 
\beq
\phi_0 \sim \varphi_{\rm max},~~M \sim \mu Q,~~\omega \sim \mu,
~~R \sim (\mu Q/\Lambda^4)^{1/3}~,
\label{qcrit}
\eeq 
where $\mu^2 \sim V(\varphi_{\rm max})/\varphi_{\rm max}^2$. Changing
$n$ from $n=4$ to $n=7$  the frequency $\omega$ ranges from $10$ MeV
to $10^{-3}$ MeV and the VEV $\phi_0$ from $10^7$ to $10^{11}$ GeV.
In notations of ref.~\cite{Gherghetta:1995dv}  Q-balls related to
$ud,~ ue,~ QL,~ ude$ flat directions are of this type. In the
following we will be interested in these Q-balls only.

Though in general the critical charge $Q_c$ depends on the
structure of the flat direction and on parameters of the underlying
unified model \cite{ru}, the exact values of the critical charge and
of $\omega$ are of little importance for us.   What is essential is
that (i)  $Q_c$ is much smaller than the baryon charge of
the neutron stars $Q_{\rm ns} \sim 10^{57}$; (ii) $\omega$ is much
smaller than the nucleon mass and smaller than the chemical potential
for the baryon number in the center of a neutron star; and (iii) 
that $\phi_0 \gg \omega$ for dark matter Q-balls. These types of
non-topological solitons  shall be dubbed CD (curved directions)
Q-balls.

Very large Q-balls can become black holes. For FD Q-balls this
happens when the global charge reaches
\beq
Q_{\rm bh} \simeq 8 \cdot 10^{61} \left(\frac{TeV}{\Lambda}\right)^4~,
\label{Q_bh_FD}
\eeq
and for CD Q-balls the corresponding number is
\beq
Q_{\rm bh} \simeq 2 \cdot 10^{53} \left(\frac{TeV}{\Lambda}\right)^2
\left(\frac{MeV}{\omega}\right)~.
\label{Q_bh_CD}
\eeq
While for FD Q-balls it exceeds the typical baryonic charge of a
star, for  CD Q-balls it is less than $Q_{\rm ns}$, which is
important for discussion below.

\begin{table}
\begin{center}
\begin{tabular}{|c|c|c|} \hline
 & FD & CD \\ \hline
$\varphi$ & $\frac 1 {\sqrt{2}} \Lambda Q^{1/4}$ 
     & $\varphi_{\rm max}$ \\ \hline
$\omega$ & $\pi \sqrt{2} \Lambda Q^{-1/4}$ 
 & $ {\Lambda^2} {\varphi^{-1}_{\rm max}}
=\pi \sqrt{2} \Lambda {Q_c}^{-1/4}=\omega_c$ 
           \\ \hline
$M$ & $4 \pi \frac {\sqrt{2}} 3  \Lambda Q^{3/4}$ & $\omega Q$ \\ \hline
$R$ & $\frac {1}{\sqrt{2}\Lambda} Q^{1/4}$ &
 ${\left(\frac 3 {8 \pi} \frac 1 {\Lambda^2 \varphi_{\rm max}} Q\right)}^{1/3} 
         =(\frac 3 2)^{1/3} {\left( Q /Q_c \right)}^{1/12} R_{FD}$
\\ \hline
\end{tabular}
\label{dks}
\caption{Summary of properties of Q-balls of different types}
\end{center}
\end{table}

A summary of the Q-ball properties can be found in Table I.

\section{Astrophysical limits}
We now ask whether such a large processing rate of baryonic matter by
the relic SUSY Q-balls could affect some astrophysical observations 
which could be used to either detect dark-matter Q-balls or rule
them out.  Q-balls are extremely dense and heavy and, therefore, they
do not stop in most astrophysical objects, such as stars or
planets~\cite{sw}.  They would stop, however, in both neutron stars
and white dwarfs.  Furthermore, simple  estimates suggest that most
neutron stars and white dwarfs are infected by Q-balls within a
short period of time~\cite{sw}.  Neutron stars and white dwarfs are,
therefore, the most reasonable places to look for Q-ball signals.

A Q-ball captured by a neutron star or a white dwarf begins consuming
baryonic matter.  At some pooint, the neutron star mass can go below its
stability limit, which is about 0.2 $M_\odot$~\cite{shapiro_teukolsky}.  Then
the neutron star explodes.   This happens when a CD Q-ball inside
becomes a black hole.  If the time scale for this process is less
than the known ages of some neutron stars and, then such Q-balls must not
exist. 
Spindown rates of some pulsars is as low as $(\dot P/P)\sim
(0.3-3)\times 10^{-10} {\rm yr}^{-1}$, which corresponds to the age in billions
of years.  
Some pulsars are also known to be (at least) as
old as 10~Gyr based on the cooling ages of their white dwarf
companions~\cite{phinney}.  One must, therefore, rule out all Q-balls that
would destroy neutrons stars in less than $10^{10}$ years. 

As we have already mentioned, the estimate of the neutron star lifetime made in
Ref.~\cite{sw} was based on an incorrect rate of the baryon number
absorption by non-topological solitons. In this section we will
revise and correct this analysis. We shall also incorporate two other effects
that were
not essential for the analysis in Ref.~\cite{sw} because of the
assumed small rate of the baryon number transfer from the neutron
matter to Q-balls. The first effect is related to the fact that the
rate of baryon number transformation on the surface of Q-balls is
large, and, therefore, there is a large pressure from the radiated pions.  
In this case the rate of absorption is determined by the
fluid dynamics and relative pressures of neutrons and pions rather 
than by the probability of matter-antimatter transformation on the Q-ball
surface.  The second effect is that the charge-energy relations are
changing after the Q-ball reaches the critical charge discussed in
Section 2, see eq. (\ref{qcrit}). Unfortunately, relatively little is
known about the internal structure of neutron stars and white
dwarfs.  Therefore, we will present two simplified models to  get an
idea of the timescale for neutron star and white dwarf consumption.

\subsection{Unitarity bound}
Let us present a lower bound on the lifetime of a neutron star infested with
Q-balls. 
Let us assume that the baryon number density near the Q-ball surface is given
by the
nuclear density $n_0 \sim 4 \cdot 10^{-3}~{\rm GeV}^3$ at the center
of the neutron star without a Q-ball  \cite{shapiro_teukolsky}. 
Then the absorption rate  simply coincides with the number of baryons
falling on the Q-ball,
\begin{equation}
\frac {dQ} {dt} \simeq \pi {r_Q}^2 n_0 v~,
\label{uni}
\end{equation}
where  $v$ is the typical nucleon velocity, taken to be of the order
of the speed of light in what follows.

Let us now assume that the nuclear density in the center of the star 
does not change with time (this is a conservative assumption because the
central density 
decreases when the baryon number of the neutron star is consumed by a
Q-ball). Then the lifetime of a star populated by FD Q-balls is of
the order of
\begin{equation}
t \sim \frac{\Lambda^2}{n_0}(Q_{\rm ns})^{1/2} \sim 10^5~
\mbox{years}~.
\label{fdq}
\end{equation}
For the CD Q-balls (a more realistic case, as the FD Q-balls are
transformed into CD Q-balls after their charge increases past $Q_c$)
the lifetime is much smaller,
\begin{equation}
t \sim \frac{(\omega_c \varphi_{\rm max}^2)^{2/3}}{n_0} 
(Q_{\rm bh})^{1/3} \sim O(1)~ \mbox{day}~.
\label{cdq}
\end{equation}

In contrast, a white dwarf is about $10^6$ times less dense than a
neutron star, so the time scale for a white dwarf to be consumed is
about $10^{11}$ years for a FD Q-ball and about $10^3$ years for a CD
Q-ball.

All but one of these estimates are  considerably smaller than the
typical ages of neutron stars or white dwarfs. We will see,
however, that the baryon number absorption is mainly determined by
the transport properties of quarks and nucleons in the interior of
the neutron stars and that the speed of baryon number consumption by
Q-balls is in fact considerably smaller. Nevertheless, it happens to
be large enough to completely exclude Q-balls (with $m=0$ in eq.
(\ref{lift})) as candidates for dark matter.

\subsection{Hydrodynamic Considerations}
It is clear that the unitarity limit is unreasonable.  However, it is
not clear what the appropriate transport mechanism inside the neutron
star will be. We know that near the Q-ball neutrons will be consumed
and turned into pions (either directly at the Q-ball surface, or in
an annihilation).  We also know  that a neutron star is held stable
by the interplay of gravity pulling inward and the tremendous
degeneracy pressure of the neutrons pressing outward. If the  change
in neutron density results in a pressure change, then the transport
will be dominated by pressure driven hydrodynamic flows.  If instead,
the  pressure remains constant, then the transport will be dominated
by either  convection or diffusion.  Therefore, we must first
determine whether or not the pressure can remain constant near the
Q-ball.

There are two simple ways to estimate the pressure near the center of
the  star.  Macroscopically, we can determine the required pressure
to maintain the star in hydrostatic equilibrium.  For a constant
density the pressure is
\beq
P=\frac 3 {8 \pi} \frac {M^2} {{m_{pl}}^2} \frac 1 {R^4}~.
\eeq
Using $M \approx 10^{57}$ GeV, and $R\approx 10 {\rm km} \approx 5 \times
10^{19} {\rm GeV}^{-1}$ yields $2 \times 10^{-4} {\rm GeV}^4 \approx 
(.12 {\rm Gev})^{4}$.

In the second approximation we use the pressure from a degenerate 
non-relativistic fermi gas.
\beq
\frac {(6 \pi^2)^{1/3}} {5 m} n^{5/3}~.
\eeq
Using the density of the star as $10^{15} \rm {g /{cm}^3}$ we find
that $P \approx 10^{-4} {\rm GeV}^4 \approx (.1 {\rm GeV})^4$.

While both approximations are highly simplified, they give the same
order of  magnitude for the pressure which we assume is reliable.  As
the Q-ball consumes baryon number, the degeneracy of neutrons will no
longer be able to maintain this high pressure.  The pressure will
then have to be maintained either thermally, or by pions before they
are able to decay into other  particles.  We will see shortly that
neither option is viable.

\subsubsection{Thermal Pressure.}

The thermal pressure from light particles is 
\beq
P \approx g \frac {T^4} {\pi^2}~,
\eeq  
where $g$ is the number of light degrees of freedom.  In our case these 
are 2 photons,  4 electrons, 2 neutrinos and possibly 3 pions (depending on
the exact temperature). This suggests that we will have an 
extrememly high temperature of about 100 MeV.

Such high temperatures can not be maintained because neutrinos carry
the  energy away faster than the Q-ball can produce it by absorbing
neutrons. Therefore we know that the pressure near the Q-ball can not
be maintained by thermal equilibrium.

\subsubsection{Pion Pressure.}
While the pressure can not be maintained by thermal effects, we may
hope that  it is maintained by the extra pions produced before they
are able to decay. Because the pions interact on strong length
scales, but decay over  electromagnetic and weak time scales, there
is a significant time when they behave much like thermal conserved
particles.  

We will therefore assume that near the Q-ball, we have a mixture of
neutrons and pions, each with an average  kinetic energy of about .1
GeV, and that the pressure of either component is simply the number
density times the kinetic energy.  The pions will  continuously
change between species, but since the average kinetic energy is lower
than their mass, no new pions will be produced.  We shall also assume
that once a pion decays, it no longer plays a significant role in
the  pressure of the system.  This is because photons, neutrinos, and
electrons all have much smaller cross sections than neutrons and
pions, and they will  immediately leave the area near the Q-ball. 
Any such particles entering the Q-ball region will have the much
lower energies typical of the colder outside region.

In this situation, the pions will cluster near the Q-ball where they
are  produced, but they will wander away from the Q-ball essentially
due to  random walk diffusion.  However, the distance they reach is
limited by their  lifetime $\tau$.  Thus, they will reach an average
distance from the Q-ball of about $\sqrt{3 \lambda \tau}$  We will
model their density profile as a  falling Gaussian with this decay
length.  A one dimensional model is  sufficient because the decay
length is much less than the radius of the Q-ball.
\beq
n_{\pi}(x)=n_{pi}(0) \exp{\left(-\frac {x^2}{3 \lambda \tau}\right)}~.
\eeq

Then the total rate of pion loss is, approximately, 
\bea
\frac {d{N_{\pi}}} {dt} & = & \frac 1 {3 \tau} 4 \pi {r_Q}^2 
\int_0^{\infty}n_{pi}(0) \exp{\left(-\frac {x^2}{\lambda \tau}\right)} 
dx \nonumber \\ 
& = & {2 \pi^{3/2}} \sqrt{\frac {\lambda} {3 \tau}} n_{\pi}(0)
{r_Q}^2~.
\eea

To maintain pressure, $n_{\pi}(0)$ needs to be about the same as the
density  of neutrons far from the Q-ball, but still near the center
of the star.  That is $n_0 \approx {.1 {\rm GeV}}^3$.  We will also
take  $\lambda \approx n_0^{-1/3}$.  That is we are letting the mean
free path be  about the spacing between particles.  Finally $\tau
\approx 10^8 {\rm GeV}^{-1}$. (This is the lifetime of a neutral pion. We
assume that the decay rate is dominated by the  electro-magnetic decays of
neutral pions.) From this we find
\beq
\frac {dN_{\pi}} {dt} = 10^{-16} Q^{2/3} {\rm GeV}~.
\eeq

This is the rate at which pions are lost to decays.  (If we require
them to be lost through weak decays, the number changes to about
$10^{-20}$.  This is  less of a change than we might expect because
the lifetime only enters as $\tau^{-1/2}$ not as $\tau^{-1}$.  In
essence, the pions are decaying slower,  but there are more of them. 
However, for such a situation to happen we must justify a lack of
neutral pions.  Given the frequent strong interactions this seems
unlikely.)  Now, one neutron can only supply about 4-5  pions, so the
rate of baryon consumption required to maintain pressure near the 
Q-ball is no more than one order of magnitude below
this.  That means
\beq
\frac {dN_{n}} {dt} \approx 10^{-17} Q^{2/3} {\rm GeV}~,
\eeq
\beq
t \approx 3 \times 10^{17} {Q_{BH}}^{1/3} {\rm GeV}^{-1} \approx 1500 \,
{\rm years}~.
\eeq

This is clearly much longer than the unitarity limit, but still well
below the known lifetimes of neutron stars.  Because they are
smaller, grow more slowly, and must consume a larger number of
neutrons, the lifetime of a  FD Q-ball is much longer, about
$10^{10}$ years.

The neutron star lifetimes are summarized in Table IV.

\begin{table}
\begin{center}
\begin{tabular} {|c|c|c|} \hline
& FD Q-balls & CD Q-balls \\ \hline
$t$ 
& $10^{10}$ years & $1500$ years \\ \hline
\end{tabular}
\label{life}
\caption{Neutron star lifetimes for Q-balls of different types.}
\end{center}
\end{table}

Of course in reality, the neutron density near the Q-ball will be
decreased slightly due to their rapid consumption.  However, as we
have shown this will  result in a drop in pressure which will cause
more neutrons to flow to the  site.  If the density dropped by
several orders of magnitude, this would  greatly increase the
lifetime of the star.  However, the pressure is so great that a
slight (order 1 or less) drop in neutron density near the Q-ball
creates a pressure gradient that is sufficiently large to replenish
the consumed  neutrons.  Such a drop would not greatly change the
lifetime of the neutron  star.

The predicted lifetime of CD Q-balls is so small that we can clearly
rule out their existence.  If they existed there would be no neutron
stars older than a  few thousand years.  FD Q-balls could still survive the
astrophysics bound. However, as mentioned earlier, large FD Q-balls are
unlikely to exist since all the flat directions are lifted for VEVs much
smaller than those for which the Q-ball reaches the size of a neutron star.  

However, this limit applies only to Q-balls with $m=0$, which  conserve baryon
number at all values of $Q$.   Q-balls for which $m \neq 0$ do not grow beyond
a certain size because the baryon number non-conserving interactions in their
interior prevent $Q$ from growing beyond the critical value $Q_c$. Such Q-balls
remain small, and, therefore, they do not consume a neutron star on
cosmologically interesting time scales. 

We note in passing that the corresponding analysis for the white dwarfs would
be more complicated because the main pressure supporting a white dwarf against
gravity comes from the Fermi pressure of degenerate electrons. This pressure is
coupled only electromagnetically to the strong dynamics of nuclei and Q-balls.
The surface temperature of the white dwarfs could be used to set a limit on the
amount of heat generated inside by the Q-ball~\cite{phinney_PC}. There are many
10~Gyr old white dwarfs that have cooled to very low temperatures and have
luminosities as low as
\begin{equation}
L_{\rm wd}=3\times 10^{-5}L_\odot = 7\times 10^{28}\, {\rm erg/ s}.
\label{Lwd}
\end{equation}
However, we do not need to consider white dwarfs in detail because, for $m=0$,
Q-balls are already ruled out by the stability of the neutron stars.  For
$m\neq 0$, the rate of energy release from consumption of nuclear matter is
much lower than the value in equation~(\ref{Lwd}). 

\section{Conclusions} 

We have revised the bounds on the relic dark-matter Q-balls based on the
existence of neutron stars and their stability on the time scale of billions of
years.  Assuming SUSY Q-balls have a sufficient abundance to be the dark
matter, relic SUSY Q-balls are captured by a neutron star shortly after its
formation.  The Q-ball consumes ambient neutrons and adds to the baryon number
of its condensate, while dissipating the excess energy in neutrinos and
photons.  Meanwhile, the neutron star mass decreases.  For most SUSY Q-balls,
this process is too rapid, and it can destabilize a neutron star on time
scales much shorter than a billion years.  Hence, such Q-balls are ruled
out.  The only SUSY Q-balls that survive this bound and remain dark-matter
candidates are those that have their flat directions lifted by the baryon
number violating operators.  Such Q-balls cannot grow beyond a certain size,
which turns out to be too small for a rapid consumption of nuclear matter.

{\bf Acknowledgments.} The authors thank E.S.~Phinney and J.~Morris for 
valuable comments.  A.K. and L.L. were supported in part by the US
Department of Energy grant DE-FG03-91ER40662 and by NASA grants
ATP02-0000-0151 and ATP03-0000-0057. The work of M.S. was supported in part 
by the Swiss Science Foundation. We thank Peter Tinyakov for helpful
discussions.



\end{document}